\documentclass[12pt]{article}
\usepackage{amssymb,amsmath,epsfig,graphicx}
\begin{document}
\title{\bf Energy Bounds in $f(R,G)$ gravity with Anisotropic Background}
\author{M. Farasat Shamir\thanks{farasat.shamir@nu.edu.pk} and
Ayesha Komal \thanks{ayeshakomal28@gmail.com}\\\\
Department of Sciences and Humanities, \\National University of
Computer and Emerging Sciences,\\ Lahore Campus, Pakistan.}
\date{}
\maketitle
\begin{abstract}
This paper investigates the energy bounds in modified Gauss-Bonnet gravity with anisotropic background. Locally rotationally symmetric Bianchi type ${I}$ cosmological model in $f(R,G)$ gravity is considered to meet this aim. Primarily, a general $f(R,G)$ model is used to develop the field equations. In this aspect, we investigate the viability of  modified gravitational theory by studying the energy conditions. We take in account four $f(R,G)$ gravity models commonly discussed in the literature. We formulate the inequalities obtained by energy conditions and investigate the viability of the above mentioned models using the Hubble, deceleration, jerk and snap parameters. Graphical analysis shows that for first two $f(R,G)$ gravity models, NEC, WEC and SEC are satisfied under suitable values of anisotropy and model parameters involved. Moreover, SEC is violated for the third and fourth models which predicts the cosmic expansion.
\end{abstract}
%{\bf PACS:} : 04.20.Jb; 98.80.-k; 98.80.Jk.

\section{Introduction}

Universe is expanding at an accelerating rate. This expansion of universe is one of the most attractive topic among the cosmologists in the recent era. It is thought that this expansion of universe is caused due to some mysterious energy with strong negative pressure. This mysterious energy is called as dark energy by the researchers. Almost 70 \% of the total universe constitutes of this dark energy. Dark energy can be described by using the Equation of State parameter \textbf{$\omega=\frac{p}{\rho}$}, where $p$ is the pressure and $\rho$ is the energy density of dark energy \cite{gar}-\cite{abi}. It is believed that modified theories can well explain the issue of dark energy. In the last decade, a lot of research work has been done in $f(R)$, $f(G)$ and $f(R,G)$ theories of gravity, where $R$ denotes the Ricci scalar and $G$ is the Guass-Bonnet invariant. Some important reviews can be helpful to understand the basics of these modified theories of gravity \cite{n22}-\cite{n44}.

Multam\"{a}ki et al. \cite{111} explored energy distributions of the Schwarzschild de-Sitter metric in $f(R)$ theory of gravity. Sotiriou and Faroni \cite{222} worked on $f(R)$ theories of gravity and discussed some important cosmological aspects.
Nojiri and Odintsov \cite{etr} designed the techniques of reconstruction for $f(G)$ gravity. They further explained that how with the application of the modified theory, the cosmological order of matter dominance deceleration-acceleration transition and acceleration phase could come forth. Houndjo et al. \cite{yui} investigated cylindrical symmetry in $f(G)$ gravity to show the existence of seven families of exact solutions. In another work \cite{uio}, cylindrical symmetry in modified field equations results the cosmic string space-time. Power law solutions with anisotropic background in $f(G)$ gravity were recently explored, and it was concluded that Bianchi type $I$ power law solutions only existed for some certain choices of $f(G)$ gravity models \cite{dfg}. Sharif and Fatima \cite{fgh} considered a viable $f(G)$ model to study non-commutative static spherically symmetric wormhole solutions in modified Guass-Bonnet gravity. Abbas et al. \cite{ghj} explored the possibility for the existence of anisotropic compact stars in $f(G)$ gravity. Wu and Ma \cite{hjk} found out exact solutions at low energy. Warm inflation in $f(G)$ theory of gravity has been discussed by Sharif and Ikram \cite{jkl}. Sharif and Fatima \cite{xcv} explored Noether symmetries by taking Friedmann-Robertson-Walker (FRW) background in $f(G)$ theory. Garcia et al. \cite{rty} considered certain choices of $f(G)$ gravity models and studied the viability of these models. The same authors \cite{cvb} considered the late-time acceleration phases of the universe.
$f(R,G)$ modified gravity has also attracted much attention of the researchers. Without involving the dark energy, the cosmic acceleration can be justified and this is the most important feature of the theory. The theory has been first introduced by Nojiri and Odintsov \cite{n11} and they showed that the theory may pass solar system tests. Linear metric perturbations have been used to describe the stability of Schwarzschild like solutions in $f(R,G)$ gravity \cite{tyu}. Shamir and Kanwal \cite{bb1} gave Noether symmetry analysis of anisotropic universe in $f(R,G)$ gravity using locally rotationally symmetric (LRS) Bianchi type $I$ universe. Shamir and Zia \cite{cc2} worked on the physical attributes of anisotropic compact stars in $f(R,G)$ gravity and concluded that the compact stars behaved usually for positive values of $f(G)$ model parameter. In a recent work, Shamir and Ahmad \cite{aa0} explored the emerging anisotropic compact stars in $f(G,T)$ gravity and gave a detailed analysis by using Karoni and Barua metric. The same authors \cite{dd3} used Noether symmetry approach to reconstruct the well-known de-Sitter solution for some specific choice of $f(G,T)$ gravity model. Capozziello et al. \cite{n77} investigated
functional form of $f(R,G)$ gravity using Noether symmetry approach. Cognola et al. \cite{n55, n66} studied modified Gauss-Bonnet
gravity to discuss dark energy and late time acceleration issues.

Some important issues in cosmology have been discussed using the energy conditions namely null energy condition (NEC), weak energy condition (WEC), strong energy condition (SEC) and dominant energy condition (DEC).
Shamir \cite{olp} explored dynamics of anisotropic power law $f(R)$ cosmology. He studied the energy conditions by considering an LRS Biachi type $I$ cosmological model and concluded that the considered model satisfy NEC, WEC and DEC for the a particular range of anisotropic parameter but SEC was not satisfied. This violation of SEC implies that the important problem of acceleration expansion of universe is supported by anisotropic universe in $f(R)$ gravity. Santos et al. \cite{ujm} studied energy conditions in $f(R)$ gravity and explored the NEC using Raychaudhri metric.
Shamir \cite{zxc} contributed by discussing the dynamics of $f(G)$ gravity in the light of energy conditions with anisotropic background. Sharif and Fatima \cite{tgb} evaluated energy conditions for Biachi type $I$ universe in $f(G)$ gravity and discussed energy conditions in modified Guass-Bonnet gravity for LRS Biachi type $I$ universe model with perfect fluid. Garcia el al. \cite{edc} evaluated energy condition in modified Guass-Bonnet gravity. They discussed the viability of $f(G)$ gravity models. Zang \cite{qaz} analyzed the violation of WEC by the inflationary Yang-Mills condensate. Shamir \cite{nbv} studied dark energy cosmological models in $f(G)$ gravity. For this purpose, LRS Biachi type $I$ cosmological model has been considered. Atazadeh and Darabi \cite{sdf} explored weak energy conditions with recent experimental values of certain cosmological parameters in $f(R,G)$ gravity. Alvarenga et al. \cite{wsx} explored testing stability of some $f(R,T)$ gravity models from energy conditions and studied the stability of de-Sitter solution. Sharif and Ikram \cite{yhn} explored energy conditions in $f(G,T)$ gravity and analyzed energy bounds. Thus it seems interesting to investigate energy condition in modified gravity.

In this article, we discuss energy conditions in $f(R,G)$ gravity with anisotropic background. LRS Bianchi type $I$ model is considered for this purpose. The article consist of two parts. In the prior part of this document, we develop gravitational field equations in $f(R,G)$ gravity. In the later half, inequalities are given which correspond to the energy conditions. The constraints resulting from the energy conditions are analyzed by considering three different models of $f(R,G)$ gravity. The last section provides the conclusion of the whole work.

\section{Einstein Field Equations}
The most general action for $f(R,G)$ gravity is given as \cite{n11}
\begin{equation}\label{1}
S=\frac{1}{2\kappa}\int d^4x\sqrt{-g}[R+f(G)]+S_{M}(g^{\mu\nu},\psi)%+T^{\mu\nu}_{mat},
\end{equation}
where $S_{M}(g^{\mu\nu},\psi)$ is the matter action, $R$ is Ricci scalar and $G$ is Guass-Bonnet invariant defined by
\begin{equation}\label{2}
G=R^{2}-4R_{\alpha\beta}R^{\alpha\beta}+R_{\alpha\beta\rho\sigma}R^{\alpha\beta\rho\sigma},
\end{equation}
where the notations $R_{\alpha\beta}$ and $R_{\alpha\beta\rho\sigma}$ are employed for the Ricci and Riemann tensors respectively. Variation of the standard action (\ref{1}) w.r.t the metric gives the following gravitational field equation
\begin{equation}\label{3}
R_{\mu\nu}-\frac{1}{2}g_{\mu\nu}R={\kappa}T^{mat}_{\mu\nu}+\Sigma_{\mu\nu},
\end{equation}
where
\begin{eqnarray}\label{4}
\Sigma_{\mu\nu}&=&\nabla_{\mu}\nabla_{\nu}f_{R}-g_{\mu\nu}\Box{f_{R}}+2R\nabla_{\mu}\nabla_{\nu}f_{G}-2g_{\mu\nu}R\Box{f_{G}}
-4R_{\mu}^{\lambda}\nabla_{\lambda}\nabla_{\nu}f_{G}\nonumber\\&&-4R_{\nu}^{\lambda}\nabla_{\lambda}\nabla_{\mu}f_{G} +4R_{\mu\nu}\Box{f{G}}+4g_{\mu\nu}R^{\alpha\beta}\nabla_{\alpha}\nabla_{\beta}f_{G}
+4R_{\mu\alpha\beta\nu}\nabla^{\alpha}\nabla^{\beta}f_{G}\nonumber\\&&-\frac{1}{2}g_{\mu\nu}(f_{R}R+f_{G}G-f(R,G))+{(1-f_{R})}{(R_{\mu\nu}-\frac{1}{2}g_{\mu\nu}R)}.
\end{eqnarray}
Here $\nabla_{\mu}$ represents the covariant derivative, $f_{R}\equiv\frac{\partial f(R,G)}{\partial{R}}$ and $f_{G}\equiv\frac{\partial f(R,G)}{\partial{G}}$ shows partial derivatives of $f(R,G)$ w.r.t $R$ and $G$ respectively.
We consider the spatially homogenous, anisotropic LRS Bianchi type $I$ metric \cite{nbv}
\begin{equation}\label{6}
d{s}^{2}=dt^{2}-L^{2}(t)d{x}^{2}-M^{2}(t)(d{y}^{2}+d{z}^{2}),
\end{equation}
where $L$ and $M$ are the cosmic scale factors. The outcome of corresponding Ricci scalar and Guass-Bonnet invariant are
 \begin{equation}\label{7}
R=-2\bigg[\frac{\ddot{L}}{L}+2\frac{\ddot{M}}{M}+2\frac{\dot{L}\dot{M}}{LM}+\frac{\dot{M^2}}{M^2}\bigg], ~~~~
%\end{equation}
%\begin{equation}\label{8}
G=8\bigg[\frac{\ddot{L}\dot{M^2}}{LM^2}+2\frac{\dot{L}\dot{M}\ddot{M}}{LM^2}\bigg],
\end{equation}
where the dot represents the time derivative. In this paper, the stress energy tensor is defined as
\begin{equation}\label{9}
T_{\nu}^{\mu}=diag[\rho,-p,-p,-p].
\end{equation}
The average scale factor $a$ and  the volume scale factor $V$ are
\begin{equation}\label{10}
a=\sqrt[3]{LM^2}, ~~~~  V=a^{3}=LM^2.
\end{equation}
We consider the Hubble parameter $H$, expansion scalar $\theta$, and shear scalar $\sigma$ in the form
\begin{equation}\nonumber\label{11}
H=\frac{1}{3}\bigg(\frac{\dot{L}}{L}+2\frac{\dot{M}}{M}\bigg),~~~~ \theta=\frac{\dot{L}}{L}+2\frac{\dot{M}}{M},~~~~ \sigma^2=\frac{1}{3}\bigg(\frac{\dot{L}}{L}-\frac{\dot{M}}{M}\bigg)^2.
\end{equation}
Using Eq. (5) the field equations (\ref{3}) take the following form
\begin{eqnarray}\label{12}
\kappa\rho=\frac{\dot{L}}{L}\dot{f_{R}}+2\frac{\dot{M}}{M}\dot{f_{R}}-12\frac{\dot{M^2}\dot{L}}{M^2L}\dot{f_{G}}
+\frac{1}{2}(f_{R}R+f_{G}G-f)+(\frac{2\dot{M}\dot{L}}{ML}+\frac{\dot{M^2}}{M^2})f_{R},
\end{eqnarray}
\begin{eqnarray}\label{13}
\kappa{p}&=&-\ddot{f_{R}}-2\frac{\ddot{M}}{M}\dot{f_{R}}+8\frac{\dot{M}\ddot{M}}{M^2}\dot{f_{G}}+4\frac{\dot{M^2}}{M^2}\ddot{f_{G}}
-\frac{1}{2}(f_{R}R+f_{G}G-f)\nonumber\\&&-(\frac{2\ddot{M}}{M}+\frac{\dot{M^2}}{M^2})f_{R},
\end{eqnarray}
\begin{eqnarray}\label{14}
\kappa{p}&=&-\ddot{f_{R}}-(\frac{\dot{L}}{L}+\frac{\dot{M}}{M})\dot{f_{R}}+(\frac{4\ddot{M}\dot{L}}{ML}
+\frac{4\ddot{L}\dot{M}}{LM})\dot{f_{G}}+4\frac{\dot{M}\dot{L}}{ML}\ddot{f_{G}}
\nonumber\\&&-\frac{1}{2}(f_{R}R+f_{G}G-f)-(\frac{\ddot{M}}{M}+\frac{\ddot{L}}{L}+\frac{\dot{M\dot{L}}}{ML})f_{R}.
\end{eqnarray}
Due to the complexity and non-linearity of the differential equations that involve five unknowns, we need some additional constraints to solve them. Direct proportionality between expansion scalar $\theta$ and shear scalar $\sigma$ can be used to give
\begin{equation}\label{15}
L=M^n,
\end{equation}
where $n$ is representing any arbitrary real number. Using Eq. (\ref{15}), the Eqs. (\ref{12})-(\ref{14}) can be written as
\begin{eqnarray}\label{16}
\kappa\rho=(\frac{n\dot{M}}{M}+2\frac{\dot{M}}{M})\dot{f_{R}}-12\frac{n\dot{M^3}}{M^{3}}\dot{f_{G}}+\frac{1}{2}(f_{R}R+f_{G}G-f)
+(\frac{2n\dot{M^2}}{M^2}+\frac{M^2}{M})f_{R},
\end{eqnarray}
\begin{eqnarray}\label{17}
\kappa{p}&=&-\ddot{f_{R}}-2\frac{\dot{M}}{M}\dot{f_{R}}+8\frac{\dot{M}\ddot{M}}{M^2}\dot{f_{G}}+4\frac{\dot{M^2}}{M^2}\ddot{f_{G}}
-\frac{1}{2}(f_{R}R+f_{G}G-f)\nonumber\\&&-\bigg(2\frac{\ddot{M}}{M}+\frac{\dot{M^2}}{M^2}\bigg)f_{R},
\end{eqnarray}
\begin{eqnarray}\label{18}
\kappa{p}=-\ddot{f_{R}}-\bigg((n+1)\frac{\dot{M}}{M}\bigg)\dot{f_{R}}+4\bigg(\frac{2n\dot{M}\ddot{M}}{M^2}
+\frac{n(n-1)\dot{M^3}}{M^3}\bigg)\dot{f_{G}}\nonumber\\+4\frac{\dot{M^2}}{M^2}\ddot{f_{G}}
-\frac{1}{2}(f_{R}R+f_{G}G-f)-\big(\frac{(n+2)\ddot{M}}{M}+\frac{n^2\dot{M^{2}}}{M^{2}}\big).
\end{eqnarray}
The Eqs. (\ref{16})-(\ref{18}) can be written in this form by using Hubble parameter, as follows
\begin{equation}\label{19}
\frac{18(2n+1)}{(n+2)^2}H^2f_{R}=2\kappa\rho-6H\dot{f_{R}}+\frac{648nH^3}{(n+2)^3}\dot{f_{G}}-f_{R}R-f_{G}G+f,
\end{equation}
\begin{eqnarray}\label{20}
\frac{3(n+3)}{(n+2)}\dot{H}f_{R}=-4\kappa(\rho+p)-2\ddot{f_{R}}+\frac{3(n+1)}{n+2}H\dot{f_{R}}+\frac{36(n+1)}{(n+2)^2}H^2\ddot{f_{G}}\nonumber\\
+\frac{72(n+1)}{(n+2)^2}H\dot{H}\dot{f_{G}}-\frac{108(n^2+n-4)}{(n+2)^3}H^3\dot{f_{G}}+\frac{9(n^2-3n-2)}{(n+2)^2}H^2f_{R}.
\end{eqnarray}
The Ricci scalar and Guass-Bonnet invariant can be rewritten in the following form
\begin{equation}\label{21}
R=-6\bigg[\frac{3(n^2+2n+3)H^2}{(n+2)^2}+\dot{H}\bigg], ~~~~
%\end{equation}
%\begin{equation}\label{22}
G=648\bigg[\frac{(n^2+2n)H^4}{(n+2)^4}+\frac{nH^2\dot{H}}{(n+2)^3}\bigg].
\end{equation}
The Eqs. (\ref{19}) and (\ref{20}) can be written in the form
\begin{equation}\label{23}
\rho_{eff}=\frac{3}{\kappa}H^2, ~~~~  p_{eff}=-\frac{1}{\kappa}(2\dot{H}+3H^2),
\end{equation}
where the effective energy density and pressure are denoted by $\rho_{eff}$ and $p_{eff}$ respectively, are as follows
\begin{eqnarray}\label{24}
\rho_{eff}=\frac{(n+2)^2}{6(2n+1)f_{R}}[2\rho-\frac{1}{\kappa}(6H\dot{f_{R}}-\frac{648n}{(n+2)^3}H^3\dot{f_{G}}+f_{R}R+f_{G}G-f)],
\end{eqnarray}
\begin{eqnarray}\label{25}
p_{eff}&=&\frac{(n+2)}{3(n+3)f_{R}}\big[-4p-\frac{1}{\kappa}(4\ddot{f_{R}}-\frac{6(n+3)}{(n+2)}H\dot{f_{R}}
+\frac{216(n^2+n+2)}{(n+2)^3}H^3\dot{f_{G}}\nonumber\\&&+\frac{144(n+1)}{(n+2)^2}H\dot{H}\dot{f_{G}}
+\frac{72(n+1)}{(n+2)^2}H^2\ddot{f_{G}}-2(f_{R}R+f_{G}G-f)
\nonumber\\&&-\frac{9(n^2+3n+2)}{(n+2)^2}H^2f_{R})\big].
\end{eqnarray}
%The above combination of equation (\ref{24})-(\ref{25}) give us the following constructive relationship.
%\begin{eqnarray}\label{26}
%\rho_{eff}+p_{eff}&=&\frac{(n+2)}{3(n+3)f_{R}}\big[4(\rho+p)+\frac{1}{\kappa}(4\ddot{f_{R}}-\frac{6(n+1)}{(n+2)}H\dot{f_{R}}
%\nonumber\\&&-\frac{216(n^2-5n+2)}{(n+2)^3}H^3\dot{f_{G}}-\frac{144(n+1)}{(n+2)^2}H\dot{H}\dot{f_{G}}
%\nonumber\\&&-\frac{72(n+1)}{(n+2)^2}H^2\ddot{f_{G}}+\frac{18(n^2-3n+2)}{(n+2)^2}H^2f_{R})\big].
%\end{eqnarray}
%(\ref{26}) that would be used in the whole paper. \\\\
Modified gravitational field equations are used to analyze the energy conditions \cite{sdf}
\begin{equation}\label{27}
NEC\Leftrightarrow\rho_{eff}+p_{eff}\geq0,
\end{equation}
\begin{equation}\label{28}
WEC\Leftrightarrow\rho_{eff}\geq0 ~~~~ \mathrm{and } ~~~~ \rho_{eff}+p_{eff}\geq0,
\end{equation}
\begin{equation}\label{29}
SEC\Leftrightarrow\rho_{eff}+3p_{eff}\geq0 ~~~~ \mathrm{and} ~~~~ \rho_{eff}+p_{eff}\geq0,
\end{equation}
\begin{equation}\label{30}
DEC\Leftrightarrow\rho_{eff}\geq0 ~~~~ \mathrm{and } ~~~~ \rho_{eff}\pm p_{eff}\geq0.
\end{equation}
The well-known definitions of deceleration, jerk and snap parameters in terms of Hubble parameter and average scale factor $a$ are
%In continuation cosmological context velocity, acceleration, snap and jerk are introduced with standard mechanics. Hence, in addition to the Hubble, deceleration, jerk and snap are denoted by
\begin{equation}\label{31}
 q=-\frac{1}{H^2}\frac{\ddot{a}}{a}, ~~   j=\frac{1}{H^3}\frac{\dot{\ddot{a}}}{a}, ~~  s=\frac{1}{H^4}\frac{\ddot{\ddot{a}}}{a},
\end{equation}
By using Eq. (\ref{31}), it follows that
\begin{equation}\label{32}
\dot{H}=-H^2(1+q), ~~ \ddot{H}=H^3(j+3q+2), ~~ \dot{\ddot{H}}=H^4(s-2j-5q-3).
\end{equation}
Now using Eq.(\ref{32}), Eqs.(\ref{24}) and (\ref{25}) take the form
\begin{eqnarray}\label{35}
\rho_{eff}&=&\frac{(n+2)^2}{6(2n+1)f_{R}}\bigg[2\rho-\frac{1}{\kappa}\big(648(\frac{(n^2+2n)}{(n+2)^4}-\frac{n(1+q)}{(n+2)^3})H^4f_{G}
\nonumber\\&&-\frac{2519424}{(n+2)^3}\big(\frac{-4(n^2+2n)(1+q)}{(n+2)^7}+\frac{2n(1+2q+q^2)}{(n+2)^6}+\frac{n(j+3q+2)}{(n+2)^6}\big)H^8f_{GG}
\nonumber\\&&+\big(23328\big(\frac{-4(n^2+2n)(1+q)}{(n+2)^4}+\frac{2n(1+2q+q^2)}{(n+2)^3}+\frac{n(j+3q+q^2)}{(n+2)^3}\big)
\nonumber\\&&-3888\big(\frac{6n(n^2+2n+3)(1+q)}{(n+2)^5}-\frac{n(j+3q+2)}{(n+2)^3}\big)\big)H^6f_{GR}
\nonumber\\&&+36\big(\frac{6(n^2+2n+3)(1+q)}{(n+2)^2}-(j+3q+2)\big)H^4f_{RR}
\nonumber\\&&-6\big(\frac{3(n^2+2n+3)}{(n+2)^2}-(1+q)\big)H^2f_{R}-f\big)\bigg],
\end{eqnarray}
\begin{eqnarray}\label{36}
p_{eff}&=&\frac{(n+2)}{3(n+3)f_{R}}\bigg[-4p+\frac{1}{\kappa}419904\big(\frac{72(n+1)H^2f_{GGG}}{(n+2)}-4f_{GGR}\big)\big(\frac{-4(n^2+2n)(1+q)}{(n+2)^4}
\nonumber\\&&+\frac{2n(1+2q+q^2)}{(n+2)^3}+\frac{n(j+3q+2)}{(n+2)^3}\big)^2H^{10}-3240\big(\frac{72(n+1)H^2f_{GG}}{(n+2)}-4f_{GR}\big)\times
\nonumber\\&&(\frac{-4(n^2+2n)(1+2q+q^2)}{(n+2)^4}+\frac{2n(1+3q+3q^2+q^3)}{(n+2)^3}+\frac{n(j+3q+2)(1+q)}{(n+2)^3})H^6
\nonumber\\&&-3888\big(\frac{144(n+1)H^2f_{GGR}}{(n+2)^2}-8f_{RRG}\big)\big(\frac{-6(n^2+2n+3)(1+q)}{(n+2)^2}+(j+3q+2)\big)\times
\nonumber\\&&\big(\frac{-4(n^2+2n)(1+q)}{(n+2)^4}+\frac{2n(1+2q+q^2)}{(n+2)^3}+\frac{n(j+3q+2)}{(n+2)^3})H^8+
\nonumber\\&&36(\frac{72(n+1)H^2f_{RRG}}{(n+2)}-4f_{RRR})(\frac{-6(n^2+2n+3)(1+q)}{(n+2)^2}+(j+3q+2)\big)^2H^6
\nonumber\\&&+648\big((\frac{216(n^2+n-4)H^3f_{GG}}{(n+2)^3}-\frac{54(n+1)H^3(1+q)f_{GG}}{(n+2)^2}-\frac{6(n+3)Hf_{GR}}{(n+2)})\times
\nonumber
\end{eqnarray}
\begin{eqnarray}
&&(\frac{-4(n^2+2n)(1+q)}{(n+2)^4}+\frac{2n(1+2q+q^2)}{(n+2)^3}+\frac{n(j+3q+2)}{(n+2)^3})\big)H^5-6\big((\frac{72(n+1)H^2f_{GR}}{(n+2)}
\nonumber\\&&-4f_{RR})(\frac{6(n^2+2n+3)(1+2q+q^2)}{(n+2)^2}+\frac{-6(n^2+2n+3)(j+3q+2)}{(n+2)^2}
\nonumber\\&&+(s-2j-5q-3))\big)H^4-6\big((\frac{216(n^2+n-4)H^3f_{GR}}{(n+2)^3}-\frac{54(n+3)H^3(1+q)f_{GR}}{(n+2)^2}
\nonumber\\&&-\frac{6(n+3)Hf_{RR}}{(n+2)})(\frac{-6(n^2+2n+3)(1+q)}{(n+2)^2}+(j+3q+2))\big)H^3
\nonumber\\&&+12\big(\frac{3(n^2+n+3)}{(n+2)^2}-1-q\big)H^2f_{R}-1296\big(\frac{n^2+2n}{(n+2)^4}-\frac{n}{(n+2)^3}\big)+2f
\nonumber\\&&+\frac{18(n^2-3n+2)H^2f_{R}}{(n+2)^2}\big)\bigg].
\end{eqnarray}
%In equation (\ref{24})-(\ref{25}) we have replaced $\dot{f_{R}}$, $\dot{f_{G}}$, $\ddot{f_{R}}$ and $\ddot{f_{G}}$ to drive the mentioned equations. Example is as below;
%\begin{equation}\label{37}
%\dot{f_{R}}=f_{RR}\dot{R}+f_{RG}\dot{G}.
%\end{equation}
Using Eqs. (\ref{35}) and (\ref{36}), we can find the respective energy conditions (See the Appendix).

\section{Graphical Analysis}

In this section, constraints caused by the energy conditions are discussed with reference to the consistency using the parameter ranges of the different $f(R,G)$ gravity models. Here we assume the vacuum case for further analysis as it has been argued that the stability of the cosmological solutions using energy conditions  can be verified by finding the accurate choices of model parameter ranges in case of vacuum \cite{mnb}. Moreover, the general results do not change by the addition of regular matter to the models because without the loss of generality the pressure and positive energy density of matter may also be added to satisfy the energy conditions \cite{sdf}.
The four well known choices of $f(R,G)$ gravity models are taken as \cite{mnb, n88}
\begin{equation}\label{42}
f_{1}(R,G)=k_{1} R^{\beta}G^{\gamma},
\end{equation}
\begin{equation}\label{43}
f_{2}(R,G)=k_{2}R+k_{3}R^{u}G^{m},
\end{equation}
\begin{equation}\label{44}
f_{3}(R,G)=k_{4}R^{u}G^{(1-m)},
\end{equation}
\begin{equation}
f_{4}(R,G)=\frac{k_{5}R^{u}}{k_{6}R^{u}+1}+k_{7}G^{m},
\end{equation}
where $k_{i}$ are arbitrary constants and $\beta$, $\gamma$, $u$, $m$ are model parameters which will gives us the information about the validity of energy conditions. Furthermore, we have used the values for the deceleration, jerk and snap parameters as $q=2$, $j=10$ and $s=-80$ respectively \cite{nbv, 333}.
\begin{figure}
\centering
\includegraphics[width=6cm,height=4cm]{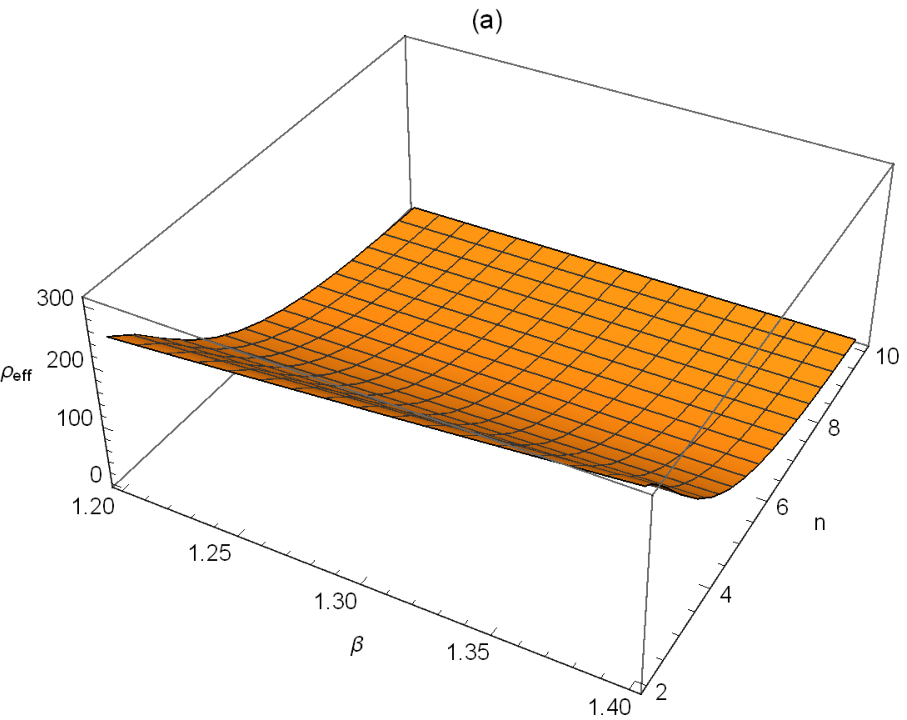}
\includegraphics[width=6cm,height=4cm]{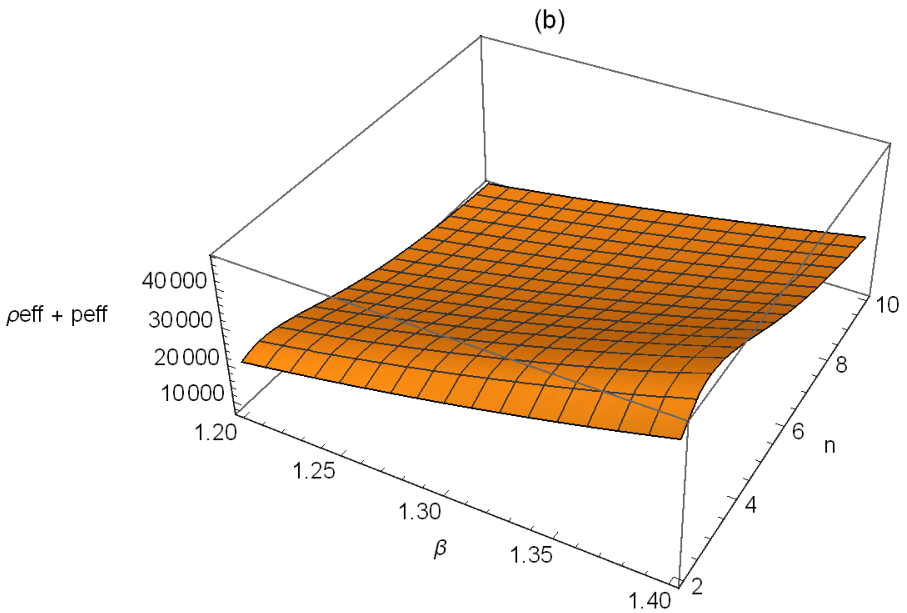}
\caption{Plots of WEC for the particular class $f_{1}(R,G)=k_1 R^{\beta}G^{\gamma}$.}
\end{figure}

\subsection{$f_{1}(R,G)=k_1 R^{\beta}G^{\gamma}$}

The inequalities $\rho_{eff}\geq0$, $\rho_{eff}\pm{p_{eff}}\geq0$ and $\rho_{eff}+3{p_{eff}}\geq0$ involving in the energy conditions are simplified by taking $\rho=0=p$. The specific values for some of the parameters are considered as the inequalities provide complex constraints. Due to this complexity, the exact analytical expressions for specific parameter range of $\beta$, $\gamma$ and $n$ can not be written here. For graphical analysis, we assign the parameter ranges for $\beta$ and $n$ and the value of $\gamma$ is fixed here $(\gamma=2)$. For model $f_{1}(R,G)$, it is clear from fig. $1(b)$ that NEC is satisfied for the given range of parameters $\beta$ and $n$. Similar trend is followed by WEC and SEC as shown in Fig. $1(a)$, $1(b)$ and Fig. $1(b)$, $2(c)$ respectively. But Fig. $2(d)$ shows violation of DEC.
%In the Fig. $1(a),~1(b)$ and $2(d)$, the plots  and  represents NEC, SEC and DEC respectively. WEC is represented by (a,b). The graphical analysis shows that NEC, WEC and SEC satisfy model $f_{1}(R,G)$ but DEC is not satisfied.
\begin{figure}
\centering
\includegraphics[width=6cm,height=4cm]{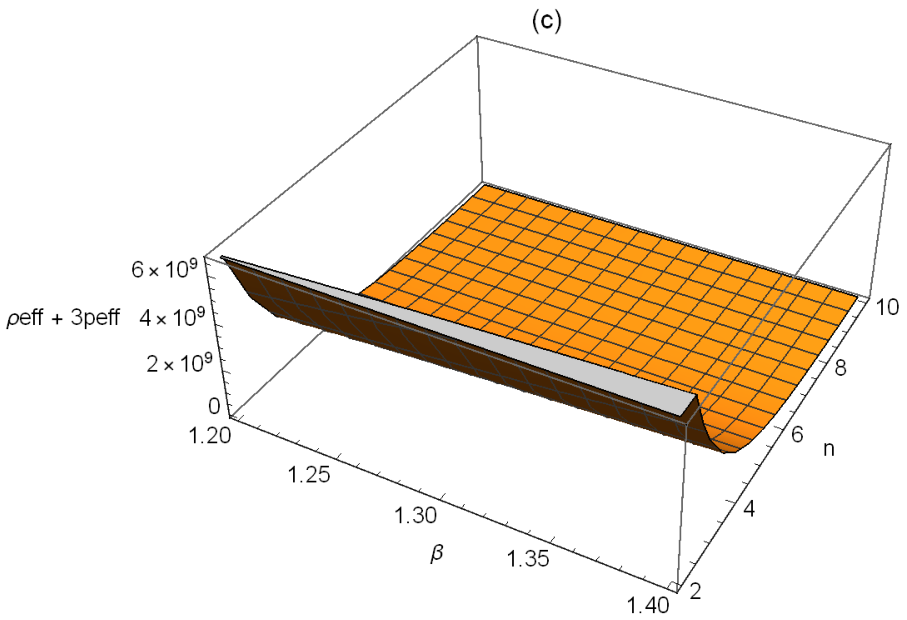}
\includegraphics[width=6cm,height=4cm]{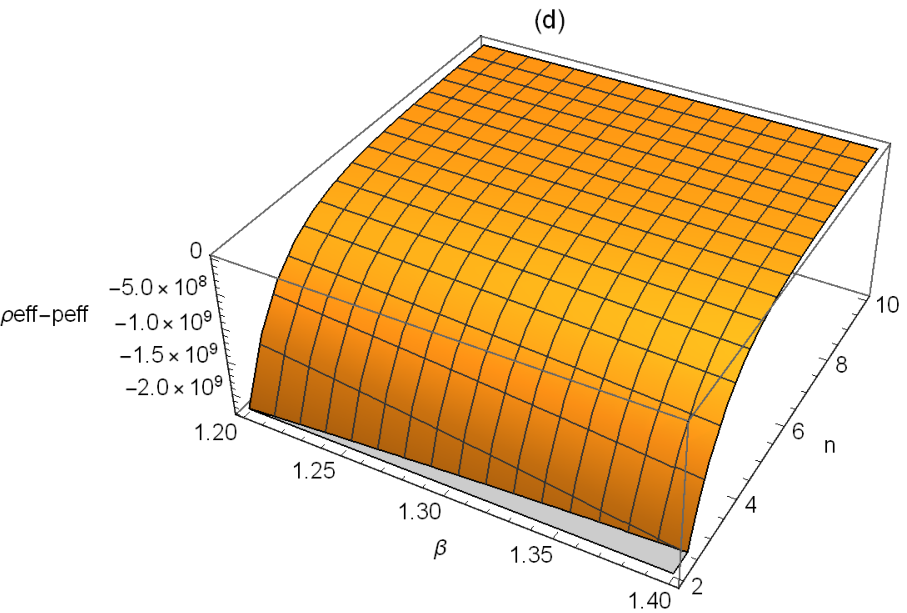}
\caption{Plots of SEC and DEC for $f_{1}(R,G)=k_1 R^{\beta}G^{\gamma}$.}
\end{figure}

\subsection{$f_{2}(R,G)=k_{2}R+k_{3}R^{u}G^{m}$}

The constants $u$ and $n$ are assigned the parameter ranges and the value of $m$ is fixed as $m=2$.  For $f_{2}(R,G)$ model, it can be seen from Fig. $3(b)$ that NEC is satisfied for the given range of parameter $u$ and $n$. Similar graphical result is shown for WEC and SEC in Figs. $3(a)$, $3(b)$ and Fig. $3(b)$, $4(c)$ respectively. But Fig. $4(d)$ shows violation of DEC.

\begin{figure}
\centering
\includegraphics[width=6cm,height=4cm]{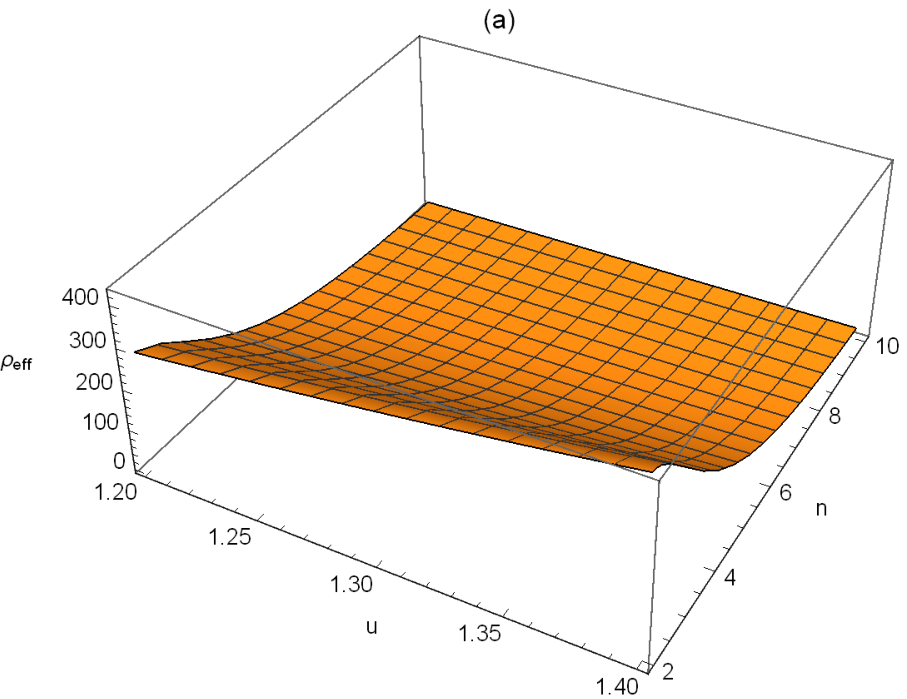}
\includegraphics[width=6cm,height=4cm]{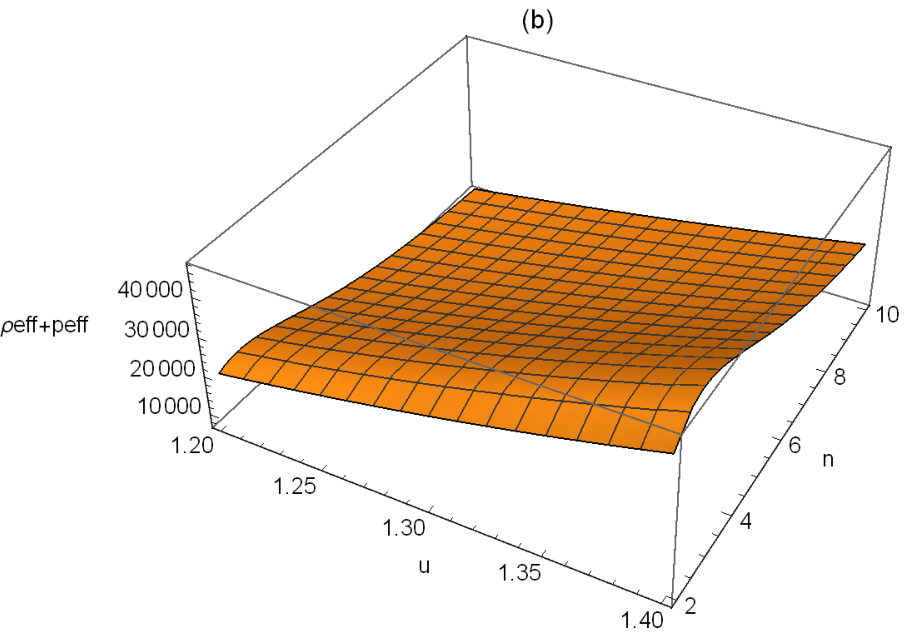}
\caption{Plots of NEC and WEC for the $f_{2}(R,G)=k_{2}R+k_{3}R^{u}G^{m}$.}
\end{figure}
\begin{figure}
\centering
\includegraphics[width=6cm,height=4cm]{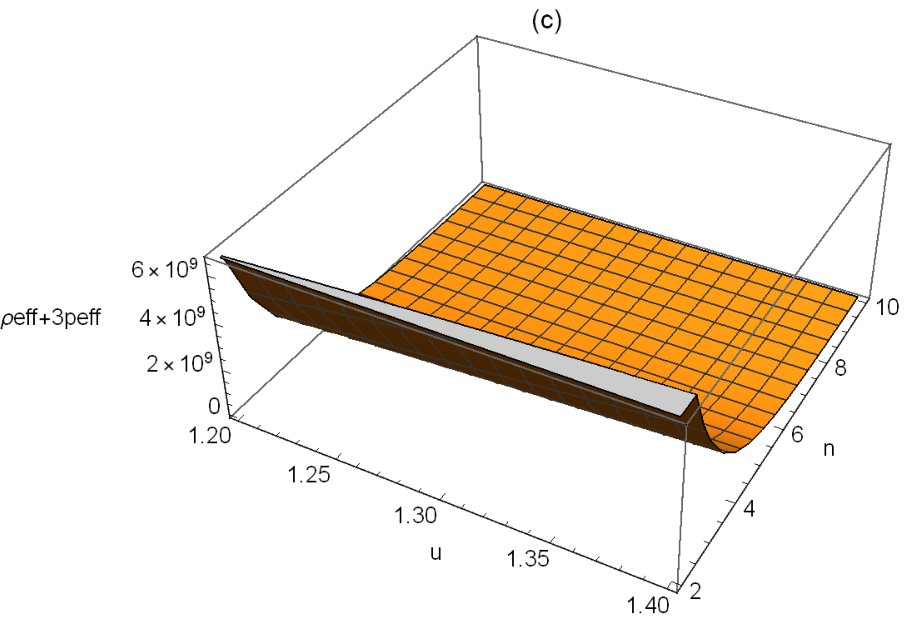}
\includegraphics[width=6cm,height=4cm]{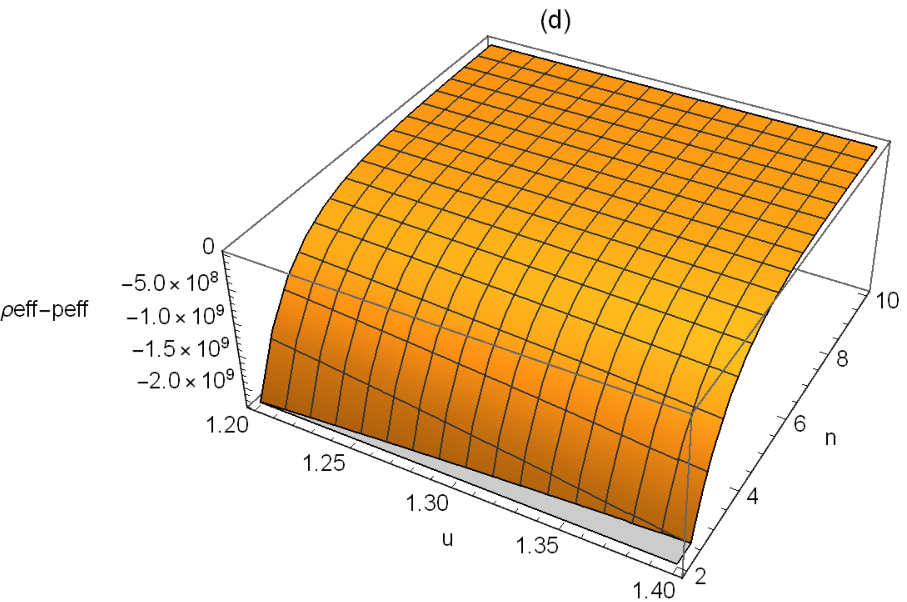}
\caption{Plots of SEC and DEC for $f_{2}(R,G)=k_{2}R+k_{3}R^{u}G^{m}$.}
\end{figure}

\subsection{$ f_{3}(R,G)=k_4R^{u}G^{1-m}$}

Here ranges are assigned to the parameters $u$ and $n$ while $m$ is fixed $(m=-1)$. The graphical analysis in Fig. $5(b)$ NEC is satisfied while WEC and DEC are also satisfied as shown in Figs. $5(a)$, $5(b)$ and Figs. $5(a)$, $5(b)$ and $6(d)$ respectively for the given range of parameters $u$ and $n$. Fig. $6(c)$ shows violation of SEC. It is interesting as it has been argued that the violation of SEC in the context of modified gravity suggests the expansion of universe \cite{Visser}.
\begin{figure}
\centering
\includegraphics[width=6cm,height=4cm]{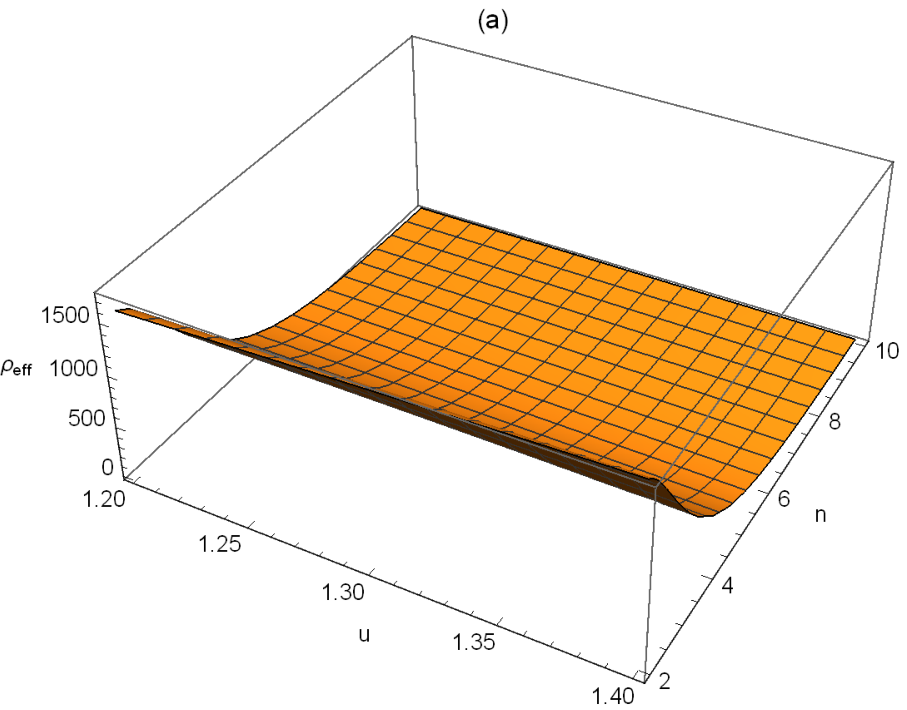}
\includegraphics[width=6cm,height=4cm]{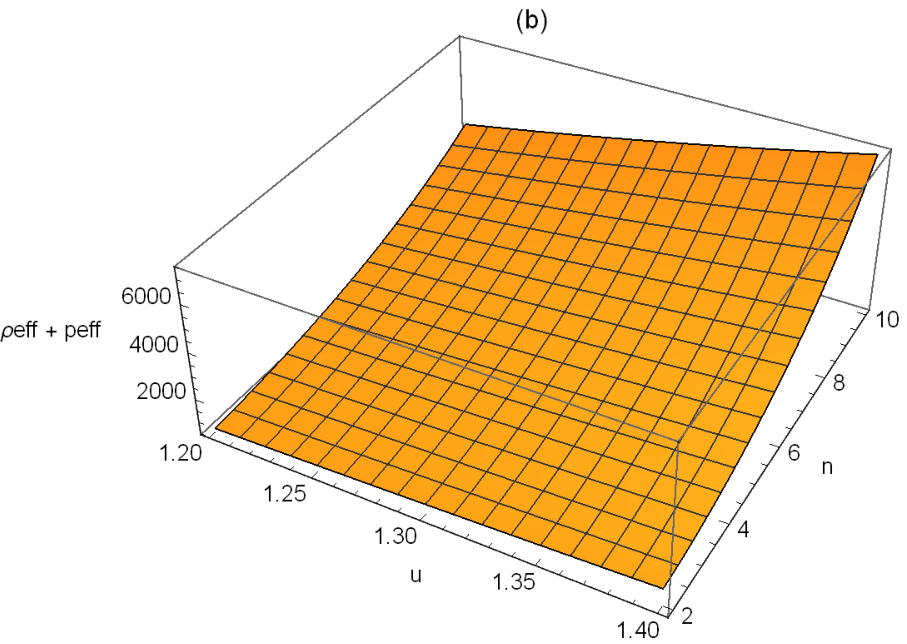}
\caption{Plots of NEC and WEC for $f_{3}(R,G)=k_4R^{u}G^{(1-m)}$.}
\end{figure}
\begin{figure}
\centering
\includegraphics[width=6cm,height=4cm]{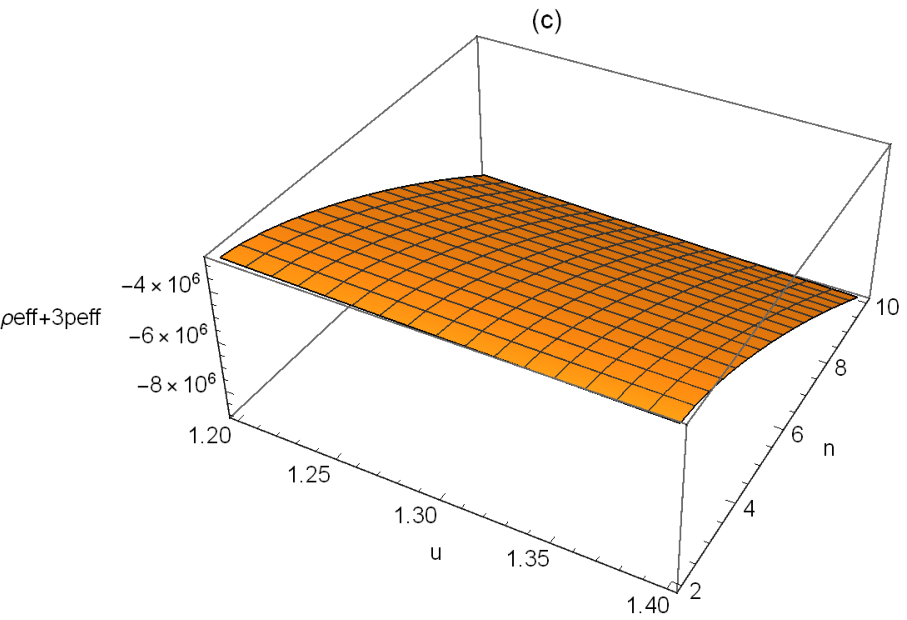}
\includegraphics[width=6cm,height=4cm]{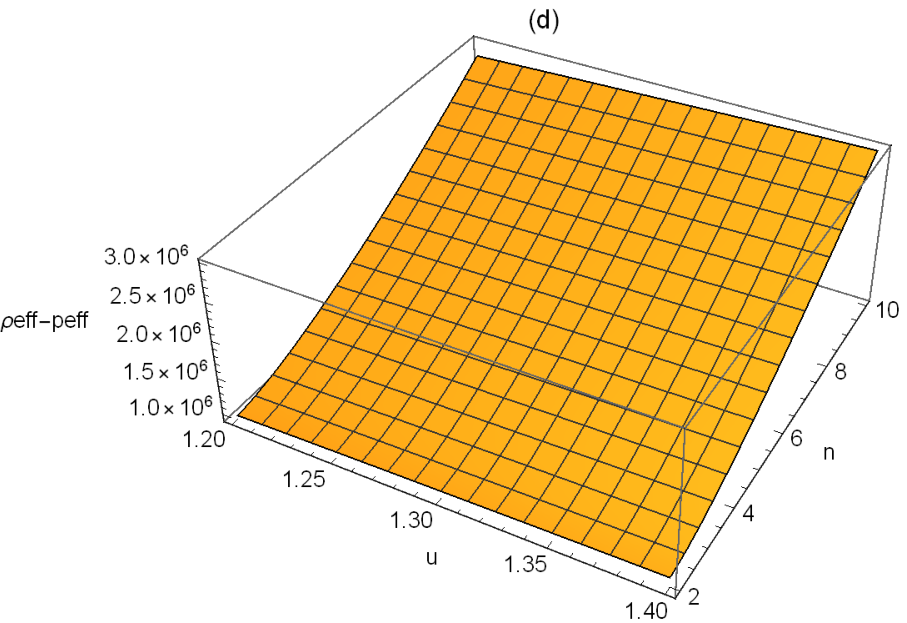}
\caption{Plots of SEC and DEC for $f_{3}(R,G)=k_4R^{u}G^{(1-m)}$.}
\end{figure}

\subsection{$f_{4}(R,G)=\frac{k_{5}R^{u}}{k_{6}R^{u}+1}+k_{7}G^{m}$}
This model is a combination of $f(R)$ and $f(G)$ models proposed in \cite{n88}. We consider $m>1$ due to the fact that late-time accelerating universe occurs if $m>1$ \cite{n88}. Here ranges are assigned to the parameters $u$ and $n$ while $m$ is fixed $(m=2)$. The graphical analysis in Figs. $(7)$ and  $8$  shows that NEC and DEC are satisfied while WEC and SEC are violated.
\begin{figure}
\centering
\includegraphics[width=6cm,height=4cm]{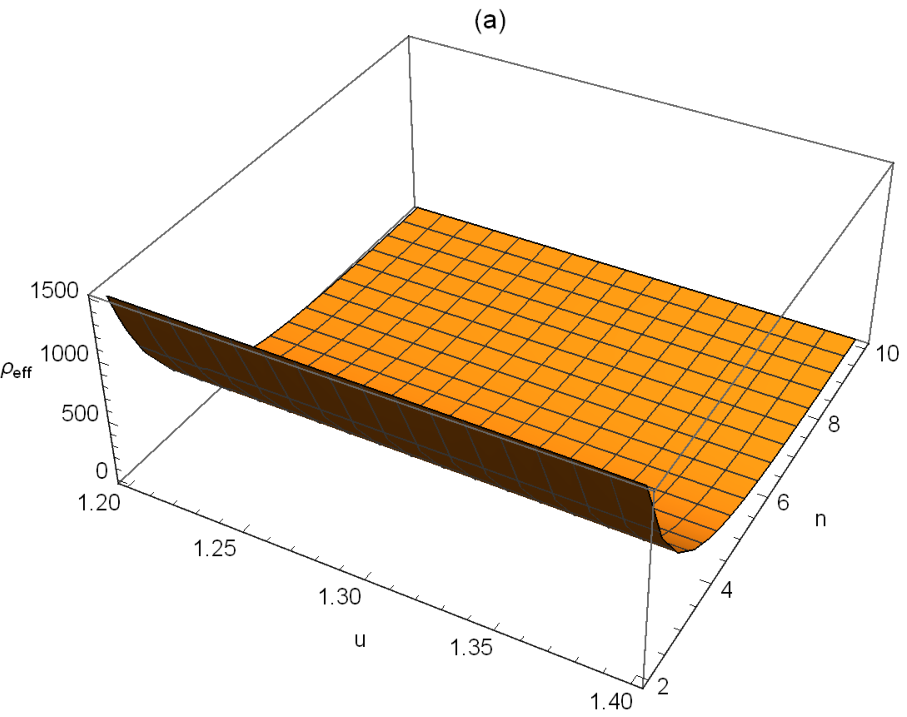}
\includegraphics[width=6cm,height=4cm]{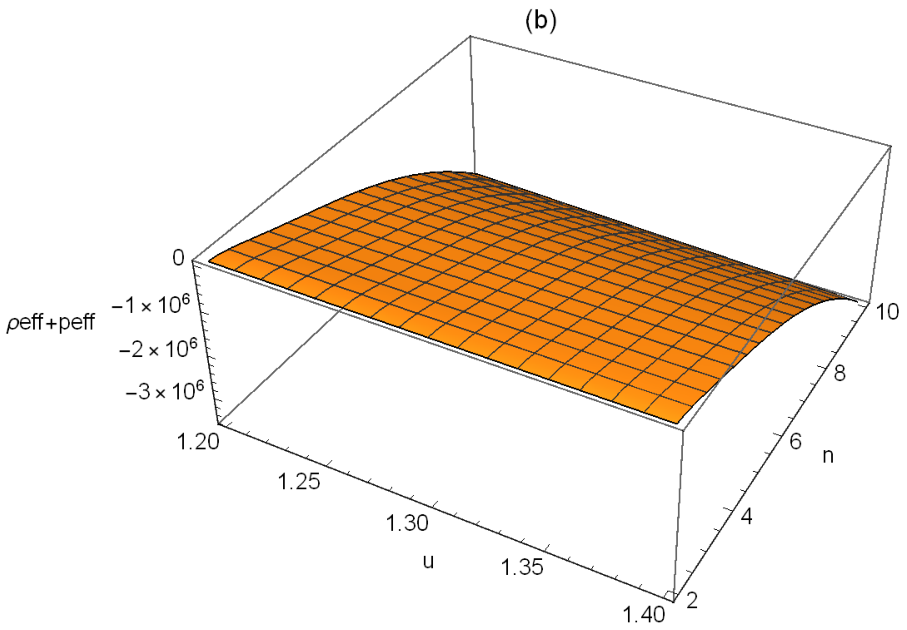}
\caption{Plots of NEC and WEC for $f_{4}(R,G)=\frac{k_{5}R^{u}}{k_{6}R^{u}+1}+k_{7}G^{m}$.}
\end{figure}
\begin{figure}
\centering
\includegraphics[width=6cm,height=4cm]{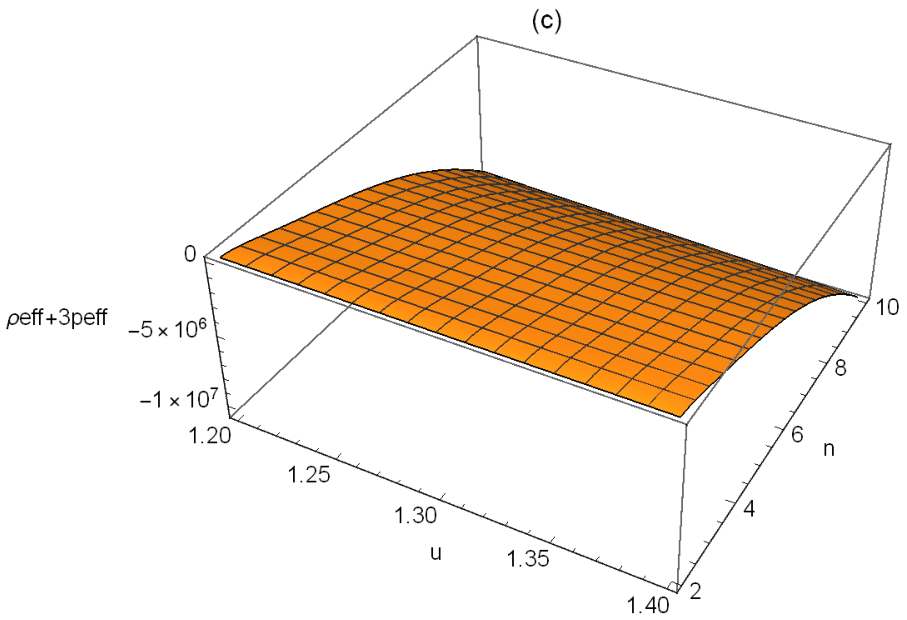}
\includegraphics[width=6cm,height=4cm]{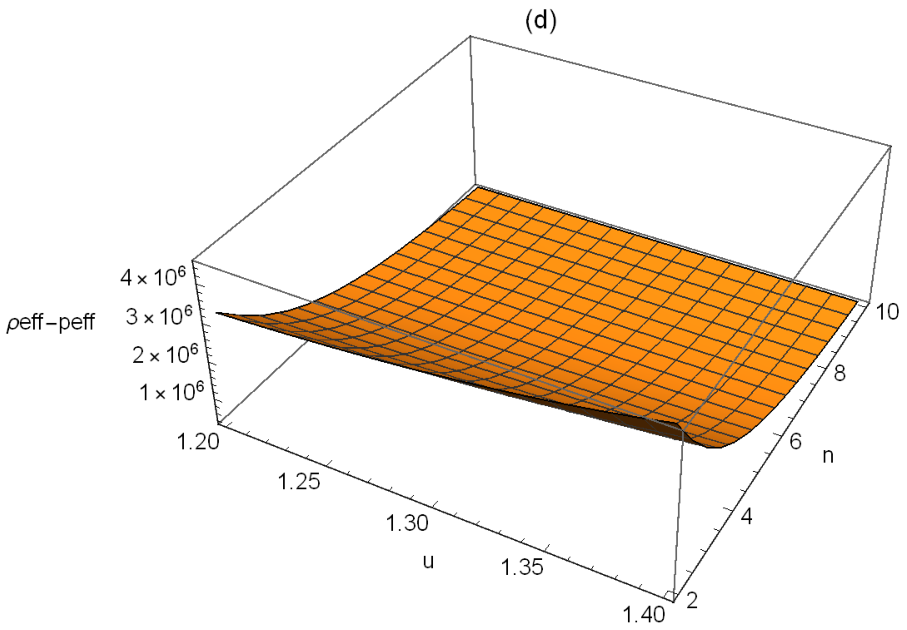}
\caption{Plots of SEC and DEC for $f_{4}(R,G)=\frac{k_{5}R^{u}}{k_{6}R^{u}+1}+k_{7}G^{m}$.}
\end{figure}

\section{Conclusion}

In this article, the $f(R,G)$ gravitational theory is discussed in the light of energy conditions. We have considered LRS Bianchi type $I$ metric for the analysis. Four realistic classes of $f(R,G)$ gravity models have been considered which are examined in the literature, accounting for the late-time cosmic acceleration and the stability of the cosmological solutions \cite{mnb}. Due to complicated and highly nonlinear nature of the field equations, we have to introduce the relation $L=M^n$ where $n$ represent the arbitrary constant and $L$, $M$ represent the metric coefficients.

We assume the vacuum case for the analysis as it has been argued that the stability of the cosmological solutions using energy conditions can be verified by finding the accurate choices of model parameter ranges in case of vacuum \cite{mnb}. Moreover, the general results do not change by the addition of regular matter to the models because without the loss of generality the pressure and positive energy density of matter may also be added to satisfy the energy conditions \cite{sdf}. Lengthy expansions that includes the inequalities can not be analyzed or handled easily. Graphical analysis shows that for $f_{1}(R,G)$ and $f_{2}(R,G)$ NEC, WEC and SEC are satisfied under suitable values of parameters involved, while DEC is violated. It is interesting to note that for $f_{3}(R,G)$ model, all conditions are satisfied except SEC. For the model $f_{4}(R,G)$, NEC and DEC are satisfied for the specific values of parameters whereas WEC and SEC are violated. It seems interesting as it has been argued that the violation of SEC in the context of modified gravity suggests the expansion of universe \cite{Visser}.

The main feature of this work is the $3$-dimensional analysis in which some model parameters are fixed and the results are examined by varying the other parameters. The work can be extended by checking the behavior of the discussed $f(R,G)$ gravity models by varying the parameters which are considered fixed in the present study.\\\\
\textbf{Acknowledgement}\\\\ We pay our sincere thanks to the honorable reviewer for the valuable comments and suggestions to improve the paper.\newpage
\textbf{Appendix}\\\\
Using Eqs. (\ref{35}) and (\ref{36}), NEC, WEC, SEC and DEC are given below respectively.\\\\
\textbf{NEC:}
\begin{eqnarray}\label{38}
&&\rho_{eff}+p_{eff}=\big[4(\rho+p)+\frac{1}{\kappa}
419904\big(\frac{-72(n+1)H^2f_{GGG}}{(n+2)}+4f_{GGR}\big)\times
\nonumber\\&&\big(\frac{-4(n^2+2n)(1+q)}{(n+2)^4}+\frac{2n(1+2q+q^2)}{(n+2)^3}+\frac{n(j+3q+2)}{(n+2)^3}\big)^2H^{10}
\nonumber\\&&-3240\big(\frac{-72(n+1)H^2f_{GG}}{(n+2)}+4f_{GR}\big)\big(\frac{-4(n^2+2n)(1+2q+q^2)}{(n+2)^4}
\nonumber\\&&+\frac{2n(1+3q+3q^2+q^3)}{(n+2)^3}+\frac{n(j+3q+2)(1+q)}{(n+2)^3}\big)H^6
\nonumber\\&&-3888\big(\frac{-144(n+1)H^2f_{GGR}}{(n+2)^2}+8f_{RRG}\big)\big(\frac{-6(n^2+2n+3)(1+q)}{(n+2)^2}
\nonumber\\&&+(j+3q+2)\big)\big(\frac{-4(n^2+2n)(1+q)}{(n+2)^4}+\frac{2n(1+2q+q^2)}{(n+2)^3}+\frac{n(j+3q+2)}{(n+2)^3}\big)H^8
\nonumber\\&&+36\big(\frac{-72(n+1)H^2f_{RRG}}{(n+2)}+4f_{RRR}\big)\big(\frac{-6(n^2+2n+3)(1+q)}{(n+2)^2}
\nonumber\\&&+(j+3q+2)\big)^2H^6+648\big(\frac{-216(n^2+n-4)H^3f_{GG}}{(n+2)^3}
\nonumber\\&&+\frac{144(n+1)H^3(1+q)f_{GG}}{(n+2)^2}-\frac{6(n+1)Hf_{GR}}{(n+2)}\big)\big(\frac{-4(n^2+2n)(1+q)}{(n+2)^4}\times
\nonumber\\&&+\frac{2n(1+2q+q^2)}{(n+2)^3}+\frac{n(j+3q+2)}{(n+2)^3}\big)H^5-6\big(\frac{-72(n+1)H^2f_{GR}}{(n+2)}+4f_{RR}\big)
\nonumber\\&&\big(\frac{6(n^2+2n+3)(1+2q+q^2)}{(n+2)^2}-\frac{6(n^2+2n+3)(j+3q+2)}{(n+2)^2}
\nonumber\\&&+(s-2j-5q-3)\big)H^4-6\big(\frac{-216(n^2+n-4)H^3f_{GR}}{(n+2)^3}
\nonumber\\&&+\frac{144(n+1)H^3(1+q)f_{GR}}{(n+2)^2}-\frac{6(n+1)}Hf_{RR}{(n+2)}\big)\big(\frac{-6(n^2+2n+3)(1+q)}{(n+2)^2}
\nonumber\\&&+(j+3q+2))H^3+\frac{18(n^2-3n+2)H^2f_{R}}{(n+2)^2}\big)\big]\geq0.
\end{eqnarray}
\textbf{WEC:}
\begin{eqnarray}\label{39}
&&\rho_{eff}=[2\rho-\frac{1}{\kappa}(648(\frac{(n^2+2n)}{(n+2)^4}-\frac{n(1+q)}{(n+2)^3})H^4f_{G}
\nonumber\\&&-\frac{2519424}{(n+2)^3}(\frac{-4(n^2+2n)(1+q)}{(n+2)^7}+\frac{2n(1+2q+q^2)}{(n+2)^6}+\frac{n(j+3q+2)}{(n+2)^6})H^8f_{GG}
\nonumber\\&&+(23328(\frac{-4(n^2+2n)(1+q)}{(n+2)^4}+\frac{2n(1+2q+q^2)}{(n+2)^3}+\frac{n(j+3q+q^2)}{(n+2)^3})
\nonumber\\&&-3888(\frac{6n(n^2+2n+3)(1+q)}{(n+2)^5}-\frac{n(j+3q+2)}{(n+2)^3}))H^6f_{GR}
\nonumber\\&&+36(\frac{6(n^2+2n+3)(1+q)}{(n+2)^2}-(j+3q+2))H^4f_{RR}
\nonumber\\&&-6(\frac{3(n^2+2n+3)}{(n+2)^2}-(1+q))H^2f_{R}-f)]\geq0,~~~~\rho_{eff}+p_{eff \geq 0}.
\end{eqnarray}
\textbf{SEC:}
\begin{eqnarray}\label{40}
&&\rho_{eff}+3p_{eff}=\frac{(n+2)^2}{6(2n+1)}\big[2\rho-\frac{1}{\kappa}\big(648(\frac{(n^2+2n)}{(n+2)^4}-\frac{n(1+q)}{(n+2)^3})H^4f_{G}
\nonumber\\&&-\frac{2519424}{(n+2)^3}\big(\frac{-4(n^2+2n)(1+q)}{(n+2)^7}+\frac{2n(1+2q+q^2)}{(n+2)^6}+\frac{n(j+3q+2)}{(n+2)^6}\big)H^8f_{GG}
\nonumber\\&&+\big(23328\big(\frac{-4(n^2+2n)(1+q)}{(n+2)^4}+\frac{2n(1+2q+q^2)}{(n+2)^3}+\frac{n(j+3q+q^2)}{(n+2)^3}\big)
\nonumber\\&&-3888\big(\frac{6n(n^2+2n+3)(1+q)}{(n+2)^5}-\frac{n(j+3q+2)}{(n+2)^3}\big)\big)H^6f_{GR}
\nonumber\\&&+36\big(\frac{6(n^2+2n+3)(1+q)}{(n+2)^2}-(j+3q+2)\big)H^4f_{RR}
\nonumber\\&&-6\big(\frac{3(n^2+2n+3)}{(n+2)^2}-(1+q)\big)H^2f_{R}-f\big)\big]
\nonumber\\&&+3\big[\frac{(n+2)}{3(n+3)}\big[-4p+\frac{1}{\kappa}419904\big(\frac{72(n+1)H^2f_{GGG}}{(n+2)}-4f_{GGR}\big)\big(\frac{-4(n^2+2n)(1+q)}{(n+2)^4}
\nonumber\\&&+\frac{2n(1+2q+q^2)}{(n+2)^3}+\frac{n(j+3q+2)}{(n+2)^3}\big)^2H^{10}-3240\big(\frac{72(n+1)H^2f_{GG}}{(n+2)}-4f_{GR}\big)\times
\nonumber
\end{eqnarray}
\begin{eqnarray}
&&(\frac{-4(n^2+2n)(1+2q+q^2)}{(n+2)^4}+\frac{2n(1+3q+3q^2+q^3)}{(n+2)^3}+\frac{n(j+3q+2)(1+q)}{(n+2)^3})H^6
\nonumber\\&&-3888\big(\frac{144(n+1)H^2f_{GGR}}{(n+2)^2}-8f_{RRG}\big)\big(\frac{-6(n^2+2n+3)(1+q)}{(n+2)^2}+(j+3q+2)\big)\times
\nonumber\\&&\big(\frac{-4(n^2+2n)(1+q)}{(n+2)^4}+\frac{2n(1+2q+q^2)}{(n+2)^3}+\frac{n(j+3q+2)}{(n+2)^3})H^8
\nonumber\\&&+36(\frac{72(n+1)H^2f_{RRG}}{(n+2)}-4f_{RRR})(\frac{-6(n^2+2n+3)(1+q)}{(n+2)^2}+(j+3q+2)\big)^2H^6
\nonumber\\&&+648\big((\frac{216(n^2+n-4)H^3f_{GG}}{(n+2)^3}-\frac{54(n+1)H^3(1+q)f_{GG}}{(n+2)^2}-\frac{6(n+3)Hf_{GR}}{(n+2)})\times
\nonumber\\&&(\frac{-4(n^2+2n)(1+q)}{(n+2)^4}+\frac{2n(1+2q+q^2)}{(n+2)^3}+\frac{n(j+3q+2)}{(n+2)^3})\big)H^5-6\big((\frac{72(n+1)H^2f_{GR}}{(n+2)}
\nonumber\\&&-4f_{RR})(\frac{6(n^2+2n+3)(1+2q+q^2)}{(n+2)^2}+\frac{-6(n^2+2n+3)(j+3q+2)}{(n+2)^2}
\nonumber\\&&+(s-2j-5q-3))\big)H^4-6\big((\frac{216(n^2+n-4)H^3f_{GR}}{(n+2)^3}-\frac{54(n+3)H^3(1+q)f_{GR}}{(n+2)^2}
\nonumber\\&&-\frac{6(n+3)Hf_{RR}}{(n+2)})(\frac{-6(n^2+2n+3)(1+q)}{(n+2)^2}+(j+3q+2))\big)H^3
\nonumber\\&&+12\big(\frac{3(n^2+n+3)}{(n+2)^2}-1-q\big)H^2f_{R}-1296\big(\frac{n^2+2n}{(n+2)^4}-\frac{n}{(n+2)^3}\big)+2f
\nonumber\\&&+\frac{18(n^2-3n+2)H^2f_{R}}{(n+2)^2}\big)\big]\big] \geq 0,~~~~\rho_{eff}+p_{eff \geq 0}.
\end{eqnarray}
\textbf{DEC:}
\begin{eqnarray}\label{41}
&&\rho_{eff}-p_{eff}=\frac{(n+2)^2}{6(2n+1)}\big[2\rho-\frac{1}{\kappa}\big(648(\frac{(n^2+2n)}{(n+2)^4}-\frac{n(1+q)}{(n+2)^3})H^4f_{G}
\nonumber\\&&-\frac{2519424}{(n+2)^3}\big(\frac{-4(n^2+2n)(1+q)}{(n+2)^7}+\frac{2n(1+2q+q^2)}{(n+2)^6}+\frac{n(j+3q+2)}{(n+2)^6}\big)H^8f_{GG}
\nonumber\\&&+\big(23328\big(\frac{-4(n^2+2n)(1+q)}{(n+2)^4}+\frac{2n(1+2q+q^2)}{(n+2)^3}+\frac{n(j+3q+q^2)}{(n+2)^3}\big)-\nonumber
\end{eqnarray}
\begin{eqnarray}
&&3888\big(\frac{6n(n^2+2n+3)(1+q)}{(n+2)^5}-\frac{n(j+3q+2)}{(n+2)^3}\big)\big)H^6f_{GR}
\nonumber\\&&+36\big(\frac{6(n^2+2n+3)(1+q)}{(n+2)^2}-(j+3q+2)\big)H^4f_{RR}
\nonumber\\&&-6\big(\frac{3(n^2+2n+3)}{(n+2)^2}-(1+q)\big)H^2f_{R}-f\big)\big]
\nonumber\\&&-\big[\frac{(n+2)}{3(n+3)}\big[-4p+\frac{1}{\kappa}419904\big(\frac{72(n+1)H^2f_{GGG}}{(n+2)}-4f_{GGR}\big)\big(\frac{-4(n^2+2n)(1+q)}{(n+2)^4}
\nonumber\\&&+\frac{2n(1+2q+q^2)}{(n+2)^3}+\frac{n(j+3q+2)}{(n+2)^3}\big)^2H^{10}-3240\big(\frac{72(n+1)H^2f_{GG}}{(n+2)}-4f_{GR}\big)\times
\nonumber\\&&(\frac{-4(n^2+2n)(1+2q+q^2)}{(n+2)^4}+\frac{2n(1+3q+3q^2+q^3)}{(n+2)^3}+\frac{n(j+3q+2)(1+q)}{(n+2)^3})H^6
\nonumber\\&&-3888\big(\frac{144(n+1)H^2f_{GGR}}{(n+2)^2}-8f_{RRG}\big)\big(\frac{-6(n^2+2n+3)(1+q)}{(n+2)^2}+(j+3q+2)\big)\times
\nonumber\\&&\big(\frac{-4(n^2+2n)(1+q)}{(n+2)^4}+\frac{2n(1+2q+q^2)}{(n+2)^3}+\frac{n(j+3q+2)}{(n+2)^3})H^8
\nonumber\\&&+36(\frac{72(n+1)H^2f_{RRG}}{(n+2)}-4f_{RRR})(\frac{-6(n^2+2n+3)(1+q)}{(n+2)^2}+(j+3q+2)\big)^2H^6
\nonumber\\&&+648\big((\frac{216(n^2+n-4)H^3f_{GG}}{(n+2)^3}-\frac{54(n+1)H^3(1+q)f_{GG}}{(n+2)^2}-\frac{6(n+3)Hf_{GR}}{(n+2)})\times
\nonumber\\&&(\frac{-4(n^2+2n)(1+q)}{(n+2)^4}+\frac{2n(1+2q+q^2)}{(n+2)^3}+\frac{n(j+3q+2)}{(n+2)^3})\big)H^5-6\big((\frac{72(n+1)H^2f_{GR}}{(n+2)}
\nonumber\\&&-4f_{RR})(\frac{6(n^2+2n+3)(1+2q+q^2)}{(n+2)^2}+\frac{-6(n^2+2n+3)(j+3q+2)}{(n+2)^2}
\nonumber\\&&+(s-2j-5q-3))\big)H^4-6\big((\frac{216(n^2+n-4)H^3f_{GR}}{(n+2)^3}-\frac{54(n+3)H^3(1+q)f_{GR}}{(n+2)^2}
\nonumber\\&&-\frac{6(n+3)Hf_{RR}}{(n+2)})(\frac{-6(n^2+2n+3)(1+q)}{(n+2)^2}+(j+3q+2))\big)H^3
\nonumber\\&&+12\big(\frac{3(n^2+n+3)}{(n+2)^2}-1-q\big)H^2f_{R}-1296\big(\frac{n^2+2n}{(n+2)^4}-\frac{n}{(n+2)^3}\big)+2f
\nonumber\\&&+\frac{18(n^2-3n+2)H^2f_{R}}{(n+2)^2}\big)\big]\big] \geq 0,~~~~\rho_{eff} \geq 0,~~~~~\rho_{eff}+p_{eff} \geq 0.
\end{eqnarray}

\end{document}